\documentclass[prb,twocolumn,superscriptaddress]{revtex4-2}

\usepackage{graphicx}
\usepackage{hyperref}
\usepackage{amsmath}
\usepackage{amssymb}
\usepackage{xcolor}
\usepackage{bm,bbm}
\usepackage[normalem]{ulem}
\usepackage{tikz}

\newenvironment{diagram}
{
\begin{tikzpicture}[baseline = (X.base),every node/.style={scale=0.7},scale=.55]
}
{
\end{tikzpicture}
}

\begin{document}

\title{A Promising Method for Strongly Correlated Electrons in Two Dimensions:\\ Gutzwiller-Guided Density Matrix Renormalization Group}

\author{Hui-Ke Jin}
\affiliation{State Key Laboratory of Quantum Functional Materials, School of Physical Science and Technology, ShanghaiTech University, Shanghai 201210, China}

\author{Rong-Yang Sun}
\affiliation{RIKEN Interdisciplinary Theoretical and Mathematical Sciences Program (iTHEMS), Wako, Saitama 351-0198, Japan}
\affiliation{Computational Materials Science Research Team, RIKEN Center for Computational Science (R-CCS), Kobe, Hyogo, 650-0047, Japan}

\author{Hong-Hao Tu}
\affiliation{Faculty of Physics and Arnold Sommerfeld Center for Theoretical Physics, Ludwig-Maximilians-Universit\"at M\"unchen, 80333 Munich, Germany}

\author{Yi Zhou}
\email{yizhou@iphy.ac.cn}
\affiliation{Institute of Physics, Chinese Academy of Sciences, Beijing 100190, China}

\date{\today}

\begin{abstract}
The study of strongly correlated electron systems remains a fundamental challenge in condensed matter physics, particularly in two-dimensional (2D) systems hosting various exotic phases of matter including quantum spin liquids, unconventional superconductivity, and topological orders. Although Density Matrix Renormalization Group (DMRG) has established itself as a pillar for simulating one-dimensional quantum systems, its application to 2D systems has long been hindered by the notorious ``local minimum'' issues.
Recent methodological breakthroughs have addressed this challenge by incorporating Gutzwiller-projected wavefunctions as initial states for DMRG simulations.
This hybrid approach, referred to as DMRG guided by Gutzwiller-projected wave functions (or Gutzwiller-guided DMRG),
has demonstrated remarkable improvements in accuracy, efficiency, and the ability to explore exotic quantum phases such as topological orders. 
This review examines the theoretical underpinnings of this approach, details key algorithmic developments, and showcases its applications in recent studies of 2D quantum systems.
\end{abstract}

\maketitle

\tableofcontents

\section{Introduction}

The study of strongly correlated electron systems has been a central theme in condensed matter physics for decades. These systems are characterized by interactions among electrons that cannot be treated in a perturbative manner, leading to emergent phenomena such as Mott insulating behavior~\cite{Anderson1973}, unconventional superconductivity~\cite{Bardeen1957, Anderson1987}, quantum spin liquids~\cite{Savary2016, Zhou2017}, and topological phases of matter~\cite{Wen02, Kitaev06}. A particularly challenging frontier lies in understanding these phenomena in two-dimensional (2D) systems, where quantum fluctuations, geometric frustration, and topology interplay to create a rich tapestry of exotic phases.

Introduced by Steven R. White in 1992~\cite{White1992}, the Density Matrix Renormalization Group (DMRG) method has emerged as a cornerstone of numerical methods for studying strongly correlated systems. Originally designed for one-dimensional (1D) systems, DMRG achieves remarkable accuracy by optimizing wave functions that take the form of matrix product states (MPSs)~\cite{Rommer1997,perezgarcia2007,Verstraete2008,Schollwock2011,XiangT-Book} Its success hinges on the area-law scaling of entanglement entropy between two subsystems~\cite{RMPAreaLaws, Hastings2007}: the entanglement entropy for gapped ground states grows proportionally to the boundary size, which remains constant in 1D systems. Thus, quantum states can be efficiently represented using MPS with a small bond dimension~\cite{RMPAreaLaws,Hastings2007}, which has established DMRG as the benchmark method for solving 1D models, such the antiferromagnetic Heisenberg chains~\cite{S1dmrg} and the Hubbard chain~\cite{White1993}.

It is natural to apply DMRG to quasi-1D systems, such as ladder and cylinder systems~\cite{Stoudenmire2012}. However, DMRG exhibits sharply distinct performance in 1D and 2D systems, which originates from the different entanglement scaling with respect to spatial dimensionality. Unlike in 1D systems, the boundary size of the subsystem in 2D systems is not constant, and thereby the entanglement entropy grows with increasing system size. 
Consequently, the bond dimension of MPS required to faithfully capture a quantum state in 2D systems can grow exponentially with the system size, rendering the traditional DMRG method increasingly computational cost~\cite{Schollwock2011, Stoudenmire2012}.  Moreover, the convergence of DMRG is not guaranteed due to the presence of the notorious ``local minima'' in the energy landscape~\cite{Jin2021}.
Despite these challenges, DMRG has been moderately successfully applied to narrow-width cylinders, providing important insights into 2D systems such as the kagome-lattice Heisenberg model~\cite{Yan2011, Depenbrock2012,He2017} and square-lattice Hubbard model~\cite{LeBlanc2015,Qin2020}.

To address the performance limitations of DMRG in two-dimensional systems, hybrid approaches have been developed to incorporate physically motivated variational Ans\"atze as initial inputs for DMRG simulations. Among these, the integration of Gutzwiller-projected wave functions with DMRG turns out to be a promising direction.
The Gutzwiller projection, first introduced by Martin Gutzwiller~\cite{Gutzwiller}, is a variational technique designed to eliminate unphysical configurations of a trial wave function, namely doubly occupied sites that are prevented by an infinitely large Hubbard $U$~\footnote{Gutzwiller's original proposal involves suppressing double occupancy using the operator $\prod_{i} \left(1-\lambda\hat{n}_{i\uparrow}\hat{n}_{i\downarrow}\right)$ with $\lambda\in[0,1]$. In this paper, we focus on the pure spin systems in which the charge degrees of freedom are completely frozen, namely, the double occupancy is fully eliminated by $\lambda=1$.}. 
Gutzwiller-projected wave functions have traditionally been combined with 
variational Monte Carlo (VMC) methods~\cite{Gros89, Ran2007}, serving as trial states to investigate quantum spin liquids and high-T$_c$ superconductors~\cite{Gros89}. 
On the other side, recent methodological progresses have shown that these trial wave functions can be formulated within the framework of tensor network states~\cite{Wu2020,Jin2020,Aghaei2020,Petrica2021,Yang2023,Li2023}. Specifically, converting them into MPS allows DMRG simulations to leverage high-quality initial states, guiding this optimization process toward a physically relevant state. This approach helps DMRG avoid local minima and accelerates convergence~\cite{Jin2021}. In addition to focusing on energetics, the quality of the trial state can be directly assessed by evaluating the wave function fidelity between itself and the DMRG-optimized state. This hybrid approach, known as ``Gutzwiller-guided DMRG'', combines the computational power of DMRG with the physical insight encoded in the variational states, emerging as a promising tool for studying 2D systems. 

Recent studies have demonstrated the potential of this hybrid method to address long-standing questions in 2D strongly correlated systems. For example, the spin-1/2 antiferromagnetic Heisenberg model on the kagome lattice, a paradigmatic system for quantum spin liquids, has been studied using this approach to reveal strong evidence for a chiral spin liquid (CSL) phase~\cite{Sun2024}. Remarkably, the boundary semion excitations in this CSL phase can be easily constructed without altering the physical Hamiltonian, highlighting the efficacy of this method in probing topological order.
Similarly, the SU(4) Kugel-Khomskii model on the triangular lattice has been identified as hosting a nematic spin-orbital liquid state with an emergent parton Fermi surface~\cite{Jin2022Bulletin}, and the same model on the honeycomb lattice is found to be a $\pi$-flux state with Dirac-type excitations~\cite{Jin2023}. 
The Kitaev honeycomb model, known for its exact solution of the entire energy spectrum, has also been investigated using this method, providing valuable insights into its topological ground-state degeneracy~\cite{Jin2021}.

In this review, we provide a comprehensive overview of the so-called ``Gutzwiller-guided DMRG'' method. We begin by discussing the theoretical framework of DMRG and Gutzwiller projection, followed by a detailed description of their integration. We then highlight applications of the method to prominent 2D models, exploring its role in uncovering exotic phases and topological orders. Finally, we discuss the challenges and future directions for this emerging approach, emphasizing its potential to revolutionize the study of 2D strongly correlated systems.

\section{Theoretical Background}
Owing to competing interactions and quantum fluctuations, strongly correlated electron systems are notoriously challenging to study using conventional techniques such as mean-field theory or perturbation theory. To effectively deal with these systems, theoretical and computational approaches usually rely on variational Ans\"atze to capture the essential entanglement features of the actual ground state for a given Hamiltonian.
In this section, we outline the theoretical framework of Gutzwiller-projected wave functions and the DMRG method, followed by a discussion of their integration within a unified hybrid framework.

\subsection{Gutzwiller-projected Wave Functions}

The Gutzwiller projection, proposed by Martin Gutzwiller in the 1960s~\cite{Gutzwiller}, provides a way to take into account correlations on top of non-interacting fermions. Broadly speaking, Gutzwiller-projected fermionic wave functions take the following form:
\begin{equation}
    |\Psi_G\rangle = P_G P_N|\Psi_0\rangle,
\end{equation}
where $|\Psi_0\rangle$ is typically chosen as a Fermi sea (Slater determinant) or BCS state (Pfaffian wave function), $P_N$ projects $|\Psi_0\rangle$ onto the subspace with fixed particle number $N$, and $P_G$ represents the projector imposing local constraints~\cite{Gros89}. For instance, the projection operator $P_G$ is defined as $P_G=\prod_i^N (1-n_{i\uparrow}n_{i\downarrow})$ to build a spin-1/2 many-body state, with $N$ denoting the total number of lattice sites.
This Gutzwiller-projected wave function is particularly effective for describing Mott insulators and quantum spin liquids, where electron-electron interactions play a dominant role. 

The Gutzwiller projector is challenging to implement analytically. One of the conventional methods for its implementation is the variational Monte Carlo (VMC) method~\cite{Gros89, Paramekanti2007, Ran2007}. A pile of previous research has established that these wave functions are particularly well-suited for capturing the physics of quantum spin liquids~\cite{Anderson1973, Savary2016}. For example, in the context of the kagome lattice Heisenberg model, Gutzwiller-projected states have been used to describe both gapless and gapped spin-liquid phases~\cite{Yan2011, Depenbrock2012}. Similarly, in doped Mott insulators, they serve as a natural framework for studying high-temperature superconductivity~\cite{Anderson1987,Anderson2004,RMP06}.

The primary advantage of Gutzwiller-projected states lies in their ability to encode strong correlations in a simple variational form. However, the optimization scheme retains a variational bias. Consequently, without complementary variational techniques, the Gutzwiller-projected wave function cannot independently confirm whether it accurately represents essential quantum fluctuations for given systems.
Furthermore, within the VMC framework, calculating key characteristic quantities, such as the entanglement spectrum and von Neumann entanglement entropy, remains computationally challenging because of inherent algorithmic limitations. To overcome these disadvantages, it remains highly desirable to extend the Gutzwiller projection by incorporating advanced numerical techniques, such as tensor networks~\cite{Wu2020, Jin2020}.
 
\subsection{Density Matrix Renormalization Group}
The DMRG method, introduced by Steven R. White in 1992~\cite{White1992}, has become the predominant numerical framework for investigating strongly correlated quantum systems. This technique can systematically approximate low-entanglement quantum states within the family of MPS~\cite{perezgarcia2007}. With the help of DMRG, the ground state of a given Hamiltonian is solved variationally through successive optimizations of the local tensor in an MPS; see details of the implementation in Ref.~\cite{Schollwock2011}.

The efficacy of DMRG for 1D systems stems from the area-law scaling of entanglement entropy~\cite{RMPAreaLaws, Hastings2007}, in which the entanglement entropy scales with the boundary area between two bipartite subsystems.
This property enables high-precision DMRG simulations with modest MPS bond dimensions, resolving seminal challenges in the Haldane chains~\cite{S1dmrg}, Hubbard chains~\cite{White1993}. In addition to static properties, DMRG can be extended to the investigation of the dynamical properties of several particular systems, such as spin chains~\cite{Hallberg1995,Kuhner1999,chebyshevmps}.

The effectiveness of DMRG in two-dimensional settings is distinctively different from its success in 1D systems, primarily due to the area-law scaling of entanglement entropy. This scaling behavior, characterized by exponentially increasing bond dimension requirements~\cite{Stoudenmire2012}, creates computational bottlenecks for simulating 2D systems. Despite these challenges, strategic implementation of DMRG on quasi-1D geometries, e.g. finite-width cylinders, has enabled groundbreaking insights into frustrated magnetism, for instance, the $S=1/2$  antiferromagnetic (AFM) Heisenberg model on the triangular and kagome lattices~\cite{White2007,Yan2011,Depenbrock2012,He2017}
Recent advances have successfully extended the capabilities of DMRG to unconventional superconducting phases in the $t-J$ model~\cite{Jiang2021,Gong2021,Lu2024,Chen2025} and exotic topological order and quantum phase transition in electronic systems~\cite{Zaletel2015,Szasz2020,Sun2021}, demonstrating its versatility beyond traditional $S=1/2$ systems.

\subsection{Integration of Gutzwiller and DMRG}

Given the complementary strengths and limitations of the Gutzwiller projection and the DMRG, a natural strategy is to combine them into a hybrid method. The idea is to construct a Gutzwiller-projected wave function in MPS form, and subsequently optimize this initial MPS using DMRG. The synergy lies in the distinct roles of two methods: Gutzwiller projection encodes essential physical insights into the initial ansatz, while DMRG provides a systematic and unbiased framework to optimize it. Clearly, the crucial aspect of this hybrid approach is the development of an efficient technique to convert a Gutzwiller-projected wave function into an MPS.

Recent methodological advances have shed light on this hybrid approach~\cite{Fishman2015,Wu2020,Jin2020, Aghaei2020,Petrica2021,Jin2022MPS,Li2025,LiuT2025}.  
For instance, Refs.~\cite{Fishman2015,Aghaei2020} introduced an efficient method for compressing a Fermi sea state into a sequence of local unitary quantum gates. Then, the MPS is created by iteratively applying these gates to a product state. In a similar way, Refs.~\cite{Wu2020,Jin2020} developed a matrix product operator-matrix product state (MPO-MPS) method with the help of maximally localized Wannier orbitals, which enables the precise and efficient representation of Fermi sea and BCS states in MPS form. Subsequent studies~\cite{Petrica2021,Jin2022MPS,LiuT2025} further enhanced these approaches with parallelization based on the Schmidt decomposition scheme.
Overall, these approaches maintain the wavefunction's entanglement structure at minimal computational cost. According to the benchmark analysis of the Kitaev honeycomb model, which can be solved exactly, DMRG optimizations with a proper initial ansatz exhibit significantly improved performance. This improvement includes avoiding local minima and targeting different topological sectors simultaneously~\cite{Jin2021}.

By combining the physical insights of Gutzwiller-projected states with the computational power of DMRG, this hybrid method has opened new avenues for exploring exotic phases, such as chiral spin liquids~\cite{Chen2021,Sun2024}, nematic spin-orbital liquids~\cite{Jin2022Bulletin}, 
and Dirac spin liquids~\cite{Jin2023,Jin2024dirac}. The method's success highlights its potential to address long-standing challenges in 2D strongly correlated systems.

\section{Methodology}

This section outlines the key computational steps of ``Gutzwiller-guided DMRG'', including the construction of Gutzwiller wave functions, conversion into MPS, and the integration into the DMRG framework.

\subsection{Construction of Gutzwiller Wave Functions}

The process begins with the construction of a suitable Gutzwiller-projected wave function. The unprojected wave function, $|\Psi_0\rangle$, is typically chosen as a mean-field solution to a quadratic Hamiltonian. Common choices for $|\Psi_0\rangle$ include BCS-like states for paired fermions, Slater determinants for free fermions, or parton wave functions for spin systems~\cite{Gros89, Wu2020}. These states encode the non-interacting physics of the system and serve as a foundation for introducing strong correlations.

Let us elaborate on this aspect using the example of $S=1/2$ spin systems with SU(2) symmetry.
To construct a fermionic mean-field theory, we first express $S=1/2$ spin operators at site $i$ in terms of two-component fermionic partons  $\psi_i = (f_{i\uparrow}, f_{i\downarrow})^T$:
\begin{equation}
\mathbf{S}_i = \frac{1}{2} \psi_i^\dagger \boldsymbol{\sigma} \psi_i,\label{eq:parton}
\end{equation}
where $\boldsymbol{\sigma}=\left(\sigma^x,\sigma^y,\sigma^z\right)$ are three Pauli matrices. This representation introduces an enlarged Hilbert space with unphysical double- and non-occupancy states. 
Nevertheless, the physical Hilbert space is restored by imposing a local single-occupancy constraint:
\begin{equation}
\psi_i^\dagger \psi_i = f_{i\uparrow}^\dagger f_{i\uparrow} + f_{i\downarrow}^\dagger f_{i\downarrow} = 1.
\end{equation}
The unphysical local states indicate a local SU(2) gauge redundancy: transformations $\psi_i \to U_i \psi_i$, $U_i \in \text{SU(2)}$, leave $\mathbf{S}_i$ invariant. The redundancy reflects the over-completeness of the parton description, necessitating projection to the physical subspace.

This parton representation allows us to systematically construct parton mean-field theory by decoupling interaction terms using Hubbard-Stratonovich fields, yielding a quadratic Hamiltonian:
\begin{equation}
H_{\text{MF}} = \sum_{ij} \left( \chi_{ij} \psi_i^\dagger \psi_j + \Delta_{ij} \psi_i^\dagger (i\sigma^y) \psi_j^\dagger + \text{h.c.} \right),
\end{equation}
where parton mean-field parameters $\chi_{ij}$ (hopping) and $\Delta_{ij}$ (singlet pairing) are also known as quantum order~\cite{Wen02}. 
The mean-field ground state of $H_{\text{MF}}$, denoted by 
$|\Phi_{\text{MF}}\rangle$ is a Slater determinant or BCS wave function, residing in the unphysical Hilbert space. To enforce the single-occupancy constraint, the Gutzwiller projection operator is applied as
\begin{equation}
|\Psi_{G}\rangle = P_G P_N |\Phi_{\text{MF}}\rangle,
\end{equation}
where, for reminder, $P_G = \prod_i (1 - n_{i\uparrow} n_{i\downarrow})$ and $P_N$ the the projector on the subspace with fixed particle number $N$.
The pivotal physics of strongly correlated systems~\cite{Anderson1973, Paramekanti2007} is captured by eliminating doubly occupied sites on top of a parton mean-field theory.

\subsection{Gutzwiller-guided DMRG}

Representing $|\Psi_G\rangle$ as an MPS is the most important step for the ``Gutzwiller-guided DMRG''. As mentioned above, several distinct algorithms have been proposed to accomplish this step~\cite{Fishman2015,Wu2020,Jin2020, Aghaei2020, Petrica2021,Jin2022MPS,Li2025,LiuT2025}.
Here, we use the brute-force MPO-MPS method~\cite{Wu2020,Jin2020} to demonstrate how the algorithm works.

The MPO-MPS method includes the following steps:
\begin{enumerate}
    \item \textbf{Identify maximally-localized Wannier orbitals}: In general, the mean-field ground state of fermionic partons can be expressed as
    \begin{equation*}
    |\Psi_{\text{MF}}\rangle=\prod_{n}d^\dagger_{n}|0\rangle.
    \end{equation*}
    Here, $d^\dagger_{n}$ is the $n$-th eigen-mode of the mean-field Hamiltonian $H_{\text{MF}}$, which should be acted into the $|0\rangle$ the vacuum of $\psi_{i}$ operators. The Pauli's exclusion principle allows us to re-express the mean-field ground state as
    \begin{equation*}
    |\Psi_{\text{MF}}\rangle=\prod_{l}\gamma^\dagger_{l}|0\rangle,
    \end{equation*}
where $\gamma^\dagger_{l}$ is the maximally-localized Wannier orbital constructed from $\{d^\dagger_{n}\}$~\cite{mlwo3,mlwo5,Wu2020,Jin2020}.
    
    \item \textbf{Represent $\gamma^\dagger_{l}$ as MPO}:
    As a single-particle mode, the maximally-localized Wannier orbital $\gamma^\dagger_{l}$ is just a superposition of parton operator $\psi_i$ and $\psi^\dagger_i$. A key observation of the MPO-MPS method is that such single-particle mode can always be represented as an MPO with bond dimension $D=2$~\cite{Wu2020,Jin2022MPS}. Without loss of generality, we take a Bogoliubov mode as an example, in which $\gamma_{l}^\dagger$ can be formally expressed as
    \begin{equation*}
    \qquad \gamma_{l}^{\dagger} =
    \left(\begin{array}{cc}
    0 & 1    \end{array}\right)\left[\prod_{\alpha}
    \left(\begin{array}{cc}
    1 & 0\\\
    V_{l\alpha}f_{\alpha}^\dagger+U_{l\alpha}f_{\alpha} & 1
    \end{array}\right)\right]
    \left(\begin{array}{cc}
    1 \\
    0
    \end{array}\right),
    \end{equation*}
    where $V_{l\alpha}$ and $U_{l\alpha}$ denotes the Wannier orbital for the particles and holes, and $\alpha=(r,s)$ is a hybrid index of lattice site $r$ and spin $s$. 
    In practice, the fermionic tensor networks (e.g, MPSs and MPOs) are usually bosonized with the Jordan-Wigner transformation of $f^{\dagger}_{\alpha} = \left[\prod_{\beta=1}^{\alpha-1}\sigma_{k}^{z}\right]\sigma^{+}_{\alpha}.$
    For instance, the corresponding MPO for the Bogoliubov mode $\gamma^{\dagger}_{l}$ can be rewritten in terms of pseudospin-1/2 as follows,
    \begin{equation*}
    \qquad \gamma^\dagger_{l} = 
    \left(0\ \ 1\right)
    \left[\prod_{\alpha}    
    \left(\begin{array}{cc}
    \sigma^0_\alpha & 0\\\    V_{l\alpha}\sigma_{\alpha}^{+}+U_{l\alpha}\sigma^{-}_{\alpha} & \sigma^{z}_{\alpha}
    \end{array}\right)\right]
    \left(\begin{array}{cc}
    1 \\
    0
    \end{array}\right),
    \end{equation*}
    where $\sigma^+$ and $\sigma^-$ denotes the raising and lowering operators for $S=1/2$ spins.

    To illustrate the next step, we would like to make use of the diagrammatic language of tensor networks. In this language, we represent tensors by geometrical shapes, where the vertical legs represent the physical degrees of freedom and the horizontal legs are the virtual degrees of freedom (the bond dimension)
    \begin{equation} 
     \qquad \gamma^\dagger_{l} = 
\begin{diagram}
\draw (1.5,1.5) circle (0.5);
\draw (1.5,1.5) node (X) {$L$};
\draw(2, 1.5) -- (2.5,1.5); 
\draw[dashed](3, 1.5) -- (4,1.5); 
\draw (4.5,1.5) -- (5,1.5);
\draw (5.5,1.5) circle (0.5);
\draw (5.5,1.5) node (X) {$\alpha$};
\draw(6, 1.5) -- (6.5,1.5); 
\draw[dashed](7, 1.5) -- (8,1.5); 
\draw (8.5,1.5) -- (9,1.5);
\draw (9.5,1.5) circle (0.5);
\draw (9.5,1.5) node (X) {$R$};
\draw (1.5,2.5) -- (1.5,2); 
\draw (1.5,1) -- (1.5,.5); 
\draw (5.5,2.5) -- (5.5,2); 
\draw (5.5,1) -- (5.5,.5); 
\draw (9.5,2.5) -- (9.5,2); 
\draw (9.5,1) -- (9.5,.5); 
\end{diagram}.
\end{equation}
Here, we list the {\em nonzero} entries of each tensor as
\begin{subequations}
\begin{equation}
\qquad \begin{diagram}
\draw (1.5,1.5) circle (0.5);
\draw (1.5,1.5) node (X) {$L$};
\draw(2, 1.5) -- (2.5,1.5); 
\draw (2.8,1.5) node (X) {$|1\rangle$};
\draw (1.5,2.5) -- (1.5,2); 
\draw (1.5,1) -- (1.5,.5); 
\end{diagram}=V_{l1}\sigma^+_{1}+U_{l1}\sigma^-_{1},
\quad
\begin{diagram}
\draw (1.5,1.5) circle (0.5);
\draw (1.5,1.5) node (X) {$L$};
\draw(2, 1.5) -- (2.5,1.5); 
\draw (2.8,1.5) node (X) {$|2\rangle$};
\draw (1.5,2.5) -- (1.5,2); 
\draw (1.5,1) -- (1.5,.5); 
\end{diagram}=\sigma^z_{1},
\end{equation}
\begin{equation}
\qquad
\begin{diagram}
\draw (0.65,1.5) node (X) {$\langle 1|$};
\draw(1, 1.5) -- (1.5,1.5); 
\draw (2, 1.5) circle (0.5);
\draw (2, 1.5) node (X) {$R$};
\draw (2,2.5) -- (2,2); 
\draw (2,1) -- (2,.5); 
\end{diagram}
=V_{lN}\sigma^+_{N}+U_{lN}\sigma^-_{N},
\quad
\begin{diagram}
\draw (0.65,1.5) node (X) {$\langle 2|$};
\draw(1, 1.5) -- (1.5,1.5); 
\draw (2, 1.5) circle (0.5);
\draw (2, 1.5) node (X) {$R$};
\draw (2,2.5) -- (2,2); 
\draw (2,1) -- (2,.5); 
\end{diagram}
=\sigma^z_N
\end{equation}
and for $1<\alpha<N$, 
\begin{equation}
\begin{split}
&\qquad \begin{diagram}
\draw (4.15,1.5) node (X) {$\langle 1|$};
\draw (4.5,1.5) -- (5,1.5);
\draw (5.5,1.5) circle (0.5);
\draw (5.5,1.5) node (X) {$\alpha$};
\draw(6, 1.5) -- (6.5,1.5); 
\draw (6.75,1.5) node (X) {$|1\rangle$};
\draw (5.5,2.5) -- (5.5,2); 
\draw (5.5,1) -- (5.5,.5); 
\end{diagram}=\sigma^0_\alpha,
\qquad
\begin{diagram}
\draw (4.15,1.5) node (X) {$\langle 2|$};
\draw (4.5,1.5) -- (5,1.5);
\draw (5.5,1.5) circle (0.5);
\draw (5.5,1.5) node (X) {$\alpha$};
\draw(6, 1.5) -- (6.5,1.5); 
\draw (6.75,1.5) node (X) {$|2\rangle$};
\draw (5.5,2.5) -- (5.5,2); 
\draw (5.5,1) -- (5.5,.5); 
\end{diagram}=\sigma^z_\alpha,\\
&\qquad \begin{diagram}
\draw (4.15,1.5) node (X) {$\langle 2|$};
\draw (4.5,1.5) -- (5,1.5);
\draw (5.5,1.5) circle (0.5);
\draw (5.5,1.5) node (X) {$\alpha$};
\draw(6, 1.5) -- (6.5,1.5); 
\draw (6.75,1.5) node (X) {$|1\rangle$};
\draw (5.5,2.5) -- (5.5,2); 
\draw (5.5,1) -- (5.5,.5); 
\end{diagram}=V_{l\alpha}\sigma^+_\alpha+U_{l\alpha}\sigma^-_\alpha.
\end{split}
\end{equation}
\end{subequations}

\item \textbf{Evolve MPS with MPO}: By noting the vacuum state is an MPS with bond dimension $D=1$, the parton mean-field state can be obtained by evolving the vacuum MPS with all of the MPOs iteratively. Each iteration step generates one new MPS, and this new MPS should be compressed by keeping a threshold bond dimension $\tilde{D}$. This can be achieved by using, e.g., the mixed canonical form of MPS with singular value decomposition.
This evolution process at the $l'$-th step can be diagrammatically represented as 
\begin{equation}
\begin{split}
&\qquad\gamma_{l'}\prod_{l<l'}\gamma_{l'}|0\rangle \\
& = \begin{diagram}
\draw  (7,-1.2) node {(compression)};
\draw (1.5,1.5) circle (0.5);
\draw (1.5,1.5) node (X) {$L$};
\draw (1.5, 0.) node (X) {$A_{l'-1}$};
\draw(2, 1.5) -- (2.5,1.5); 
\draw[rounded corners] (1,0.5) rectangle (2,-0.5);
\draw(2, 0.) -- (2.5,0.); 
\draw[dashed](3, 1.5) -- (4,1.5); 
\draw[dashed](3, 0.) -- (4,0.); 
\draw (4.5,1.5) -- (5,1.5);
\draw (4.5,0.) -- (5,0.);
\draw (5.5,1.5) circle (0.5);
\draw[rounded corners] (5,0.5) rectangle (6,-0.5);
\draw (5.5,1.5) node (X) {$\alpha$};
\draw (5.5, 0.) node (X) {$A_{l'-1}$};
\draw(6, 1.5) -- (6.5,1.5); 
\draw(6, 0.) -- (6.5,0.); 
\draw[dashed](7, 1.5) -- (8,1.5); 
\draw[dashed](7, 0.) -- (8,0.); 
\draw (8.5,1.5) -- (9,1.5);
\draw (8.5,0.) -- (9,0.);
\draw (9.5,1.5) circle (0.5);
\draw[rounded corners] (9,0.5) rectangle (10,-0.5);
\draw (9.5,1.5) node (X) {$R$};
\draw (9.5, 0.) node (X) {$A_{l'-1}$};
\draw (1.5,2.5) -- (1.5,2); 
\draw (1.5,1) -- (1.5,.5); 
\draw (5.5,2.5) -- (5.5,2); 
\draw (5.5,1) -- (5.5,.5); 
\draw (9.5,2.5) -- (9.5,2); 
\draw (9.5,1) -- (9.5,.5); 
\draw[<-,double]  (5.5,-1.8) -- (5.5,-0.8);
\end{diagram}\\
& \approx \begin{diagram}
\draw(2, 1) -- (2.5,1); 
\draw[rounded corners] (1, 1.5) rectangle (2,0.5);
\draw (1.5, 1.) node (X) {$A_{l'}$};
\draw[dashed](3, 1.) -- (4,1.); 
\draw (4.5,1.) -- (5,1.);
\draw[rounded corners] (5,1.5) rectangle (6,0.5);
\draw(6, 1.) -- (6.5,1.); 
\draw (5.5, 1.) node (X) {$A_{l'}$};
\draw[dashed](7, 1.) -- (8,1.); 
\draw (8.5,1.) -- (9,1.);
\draw[rounded corners] (9,1.5) rectangle (10,0.5);
\draw (9.5, 1.) node (X) {$A_{l'}$};
\draw (1.5,2) -- (1.5,1.5); 
\draw (5.5,2) -- (5.5,1.5); 
\draw (9.5,2) -- (9.5,1.5); 
\end{diagram},
\end{split}    
\end{equation}
where we use $A_{l'}$ to denote the intermediate MPS with a number of $l'$ Bogoliubov modes occupied.

    \item \textbf{Apply Gutzwiller projector}: The Gutzwiller projector is a product of local operators, which is just an MPO with bond dimension $D=1$. Therefore, the Gutzwiller projection can be easily implemented by updating (often simply deleting) each local tensor of MPS.
\end{enumerate}

The compression step is critical for managing computational resources while preserving the entanglement structure of the wave function. 
An important aspect of the MPO-MPS approach is the monitoring of the MPO-MPS compression precision. 
In order to quantify the accuracy of the final MPS, the accumulated truncation errorc is introduced as
\begin{align}
    &\epsilon_{\mathrm{trunc}}(\tilde{D})=1-\prod_{m=1}^{2N}F^{(m)}(\tilde{D}), \\
    &F^{(m)}(\tilde{D})=1-\sum^{2N}_{j=1}\epsilon^{(m)}_{j}(\tilde{D}).
 \label{eq:trunc_error}
\end{align}
Here, $\epsilon^{(m)}_{j}(\tilde{D})$ is the sum of the discarded squared singular values at the $j$-th bond of the $m$-th MPO-evolved MPS. Notice that $F^{(m)}(\tilde{D})$ is a lower bound of the overlap between MPO-evolved MPS and truncated MPS in the $m$-th MPO-MPS step.

Once $|\Psi_G\rangle$ has been converted into an MPS, it can serve as a physically motivated initial ansatz for DMRG computations. The optimization proceeds through iterative tensor updates which minimize the variational energy:
\begin{equation}
E = \langle\Psi|H|\Psi\rangle/\langle\Psi|\Psi\rangle.
\end{equation}
During each DMRG sweep, the algorithm employs an MPO representation of the given Hamiltonian while solving the eigenvalue problems for each individual MPS tensor, in which the bond dimension of the MPS is gradually increased from $\tilde{D}$ to $D$.
As we will clarify below, the initial state provided by $|\Psi_G\rangle$ significantly reduces the risk that the DMRG gets stuck in unphysical local minima, which is a common challenge for systems with complex energy landscapes~\cite{Stoudenmire2012,Jin2021}.

The fermionic parton representation in Eq.~\eqref{eq:parton} can be naturally extended to a bosonic counterpart, such as the well-known Schwinger boson representation for $S=1/2$ spins. Utilizing this framework, one can develop a Schwinger-boson mean-field theory as an effective description for $S=1/2$ systems. Within this approach, a Gutzwiller projected trial wavefunction, analogous to that in the fermionic case, can be systematically constructed. 
By contrast, the elementary quasiparticles in such a parton state are bosonic, which implies two key differences: i) Before Gutzwiller projection, the local Hilbert space is not finite-dimensional. ii) There are no “Wannier orbitals” for bosons.
Due to these reasons, the fidelity of converting a Gutzwiller projected bosonic state into an MPS is not comparable to that of converting a Gutzwiller projected fermionic Gaussian state into MPS. 
Although an algorithm for bosonic wave functions has been proposed in Ref.~\cite{Wu2020}, the development of a more efficient algorithm for converting bosonic Gaussian states into MPS is highly demanded.

\section{Applications to 2D Systems}

By leveraging physically motivated Gutzwiller-projected wave functions as initial ansatz, this hybrid method addresses key challenges in 2D strongly correlated systems, as evidenced by the improved computational efficiency and precision by orders of magnitude, see Table.~\ref{tab:Performance}.
In this section, we discuss its applications to several famous 2D systems, illustrating its effectiveness in uncovering ground-state properties, quasiparticle excitations, and topological order.

\begin{table}
\caption{ 
The relative running wall time of Gutzwiller-guided DMRG compared to conventional DMRG for various models presented in this paper. }\label{tab:Performance}
    \begin{tabular}{  p{4.6cm} p{3.4cm} }
		\hline
		\hline
		\rule{0pt}{2.5ex}    
          Models & Running Time Ratio \\ 
                & (Gutz-DMRG/DMRG) \\
		\hline
       Kitaev honeycomb model  &  0.39 \\  
        \hline      
       Triangular-lattice SU(4)  & \\ 
       Kugel-Khomskii model &  0.48 \\
       \hline
       $S=1/2$ kagome-lattice   & \\
       AFM Heisenberg model   &  0.62  \\
		\hline
		\hline
    \end{tabular}
\end{table}

\subsection{Kitaev Honeycomb Model}

\begin{figure}
    \centering
    \includegraphics[width=1.0\linewidth]{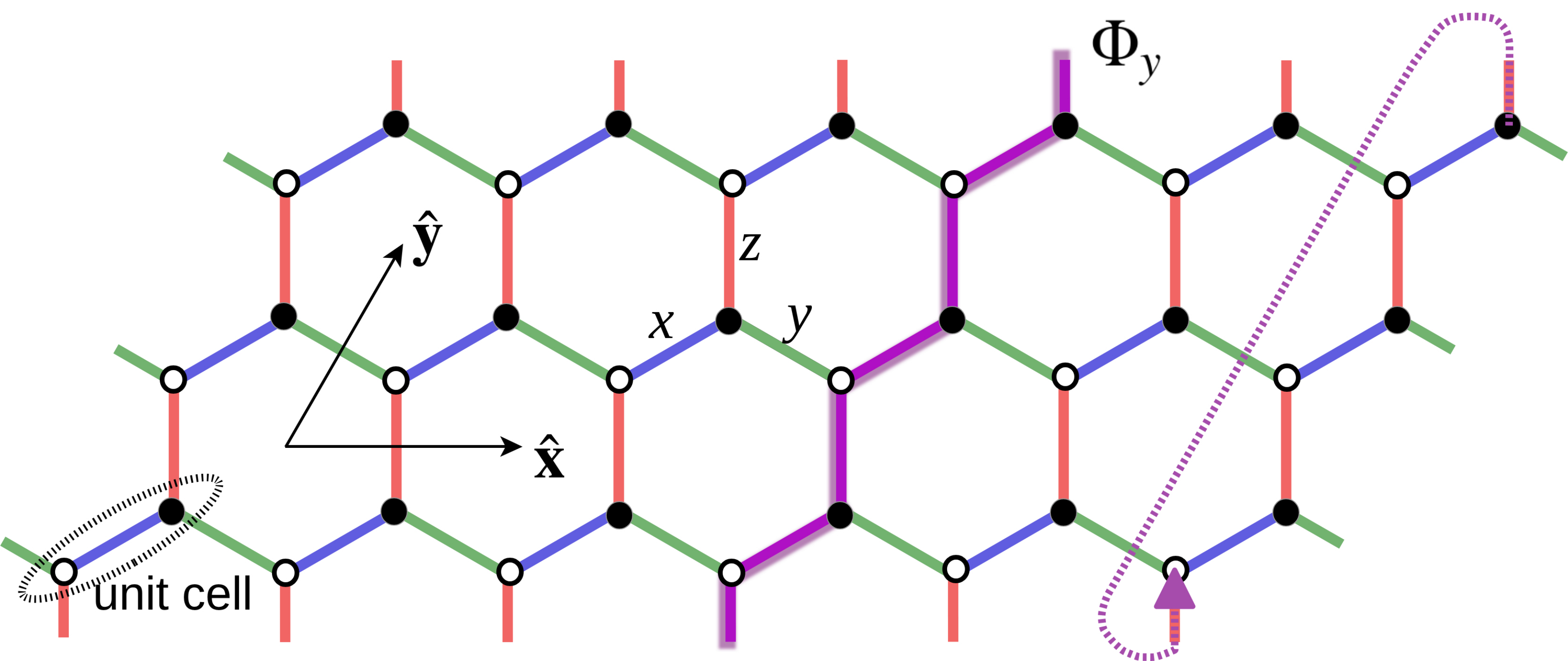}
    \caption{Kitaev honeycomb model on a cylindrical geometry, in which the $x$-boundary is open while the $y$-boundary remains periodic. Black dots and white circles stand for $A$ and $B$ sublattices. The $a$-type ($a=x,y,z$) Ising interactions are defined on the $a$ type of nearest-neighbor bonds.  The purple zigzag line indicates a closed loop $\mathcal{C}$  along which the Wilson loop operator $Wy$ is defined, see Eq.~\eqref{eq:Wy}.
    Figure from Ref.~\cite{Jin2021}.}
    \label{fig:KitaevMolde}
\end{figure}

The $S=1/2$ Kitaev honeycomb model~\cite{Kitaev06} is a paradigmatic system in the study of quantum spin liquids, offering an exactly solvable framework for exploring fractionalization and topological ground-state degeneracy. The Hamiltonian of this model is given by
\begin{equation}
H_{\text{Kitaev}} = J_x \sum_{\langle ij \rangle_x} \sigma_i^x \sigma_j^x +J_y \sum_{\langle ij \rangle_y} \sigma_i^y \sigma_j^y +J_z \sum_{\langle ij \rangle_z} \sigma_i^z \sigma_j^z,
\end{equation}
where $\sigma_i^a$ ($a = x, y, z$) are three Pauli matrices, and $\langle ij \rangle_a$ denotes nearest-neighbor bonds of type $a$; see Fig.~\ref{fig:KitaevMolde}. The bond-dependent anisotropic interactions lead to a highly frustrated system, where the ground state is known to be a Kitaev spin liquid with emergent Majorana fermions coupled to a static $\mathbb{Z}_2$ gauge field.

The exact solvability of the Kitaev honeycomb model provides an ideal benchmark for testing and validating the Gutzwiller-guided DMRG method.
This observation is clearly illustrated by using Kitaev's four-Majorana representation of spin as $\sigma^{a}_j=ic^{a}_jc^{0}_j$, where $c^a$ and $c^0$ are called gauge and itinerant Majorana fermions, respectively. Note that this representation is equivalent to the parton representation in Eq.~\eqref{eq:parton}, up to an SU(2) gauge transformation. Nevertheless, the local single-occupancy constraint becomes $D_j\equiv c^{x}_jc^{y}_jc^{z}_jc^0_j=1$, and thereby the Gutzwiller projector for restoring the physical Hilbert space reads $P_G = \prod_j \frac{1+D_j}{2}$. 
Under this representation, $H_{\text{Kitaev}}$ becomes
\begin{align}
    H_{\text{Kitaev}} = -i \sum_a\sum_{\langle jk \rangle\in a}J_a{u}_{jk} c^0_j{}c^0_k,
    \label{eq:Heff}
\end{align}
where the static $\mathbb{Z}_2$ gauge field $u_{jk} \equiv ic_{j}^{a}c^{a}_{k}$ lives on an $a$-type bond. Noting that this gauge field commutes with the Hamiltonian for all different bonds, one can replace the operators of $u_{jk}$ by their eigenvalues $\pm 1$. After fixing the static $\mathbb{Z}_2$ gauge field, $H_{\text{Kitaev}}$ becomes a quadratic Hamiltonian of itinerant Majorana fermions, whose ground state is a BCS wave function. 

\subsubsection{Benchmark results}

As a Gutzwiller-projected BCS wave function, the ground state of the Kitaev honeycomb model can be converted into an MPS using the MPO-MPS method~\cite{Jin2021}. On a finite cylinder with $L_y\times{}L_x=4\times10$, the MPO-MPS method leads to a tiny accumulated truncation error of $\epsilon_{\text{trunc}}\sim{}10^{-2}$ with a small bond dimension $\tilde{D}=200$.

We further perform DMRG optimization with the MPSs prepared from Gutzwiller-projected states, gradually increasing the bond dimension from $\tilde{D}$ to $D$. For comparison, we also generate an MPS with random tensors with bond dimension $\tilde{D}$, and then optimize it with the same DMRG scheme until a converged MPS of bond dimension $D$ is obtained. 
The calculations are performed on an $L_x \times L_y =6 \times 6$ cylinder. The truncation errors of DMRG are kept to smaller than $10^{-9}$ during the optimization of DMRG. For comparison, we calculate the relative energy deviation $\delta E_g$ between the DMRG energy and exact ground-state energy.
We find that 1) the converged MPS reveals that Gutzwiller Ans\"atze achieve the optimal result with a final bond dimension of $D=6500$, notably smaller than the required $D=8000$ for random MPS initialization; 2) the required number of sweeps for Gutzwiller Ans\"atze is roughly half of those for random MPSs; 3) the relative energy deviation $\delta E_g$ for the Gutzwiller Ans\"atze is two orders of magnitude smaller than that of a random MPS. In general, this benchmark study fully demonstrates the superiority of Gutzwiller-guided DMRG.

\subsubsection{Distinguishing topological sectors}
The Kitaev honeycomb model exhibits topological degeneracy in manifolds with non-trivial topology, such as cylinders or tori. Here, we adopt cylindrical boundary conditions, where the honeycomb lattice is embedded on a finite cylinder with $L_x$ ($L_y$) unit cells along the open (periodic) directions; see Fig.~\ref{fig:KitaevMolde}. The Hamiltonian $H_{\text{Kitaev}}$ now commutes with the Wilson loop operator wrapping around the cylinder, e.g., 
\begin{equation}
W_y = -\prod_{j \in \mathcal{C}} \sigma^y_j\label{eq:Wy}
\end{equation} 
with $\mathcal{C}$ being a closed loop as shown in Fig.~\ref{fig:KitaevMolde}. The eigenvalue of $W_y$ is just the product of the static $\mathbb{Z}_2$ gauge fields along the loop,
\begin{align}
    W_y\vert \Psi_G\rangle = \Phi_y\vert\Psi_G\rangle
    \label{eq:wilson-loop}
\end{align}
with $\Phi_y=\prod_{\langle jk \rangle \in \mathcal{C}} u_{jk} = \pm 1$ labeling two different topological ground-state sectors.

The Gutzwiller-guided DMRG method is an ideal tool to target ground states in distinct topological sectors~\cite{Jin2021}. Moreover, during the DMRG sweeps initialized with Gutzwiller-projected states, the eigenvalue of the Wilson loop operator ($\Phi_y=\pm 1$) is preserved, i.e., the MPS stays in the respective sector. This is very useful for studying topologically ordered states with (quasi-)degenerate ground states on cylinders.
In contrast, we found that the DMRG initialized with a random MPS always converges to an MPS in $\Phi_y=-1$ sector on a $6 \times 6$ cylinder. However, exact results indicate that for a finite cylinder, the ground-state energy in the $\Phi_y=-1$ sector is higher than that in the $\Phi_y=1$ sector. For instance, the energy difference on the $6 \times 6$ cylinder is given by $E_g(\Phi_y=-1)-E_g(\Phi_y=1)\approx 0.084$. 

The above result implies that the DMRG with a random initial ansatz gets stuck in a local minimum. On contrast, when the initial state already captures key features of the expected ground state, e.g., topological properties or entanglement structures, the DMRG optimization is guided toward a more relevant region of the energy landscape, reducing the likelihood of converging to unphysical local minima. It is noted that any initial state introduces bias, including the random MPS, which could potentially limit the exploration of the full Hilbert space. However, as demonstrated in this example, in the context of DMRG for 2D systems, this controlled bias is advantageous, as it leverages prior knowledge to enhance efficiency and accuracy, much like informed initial guesses in other variational methods (e.g., in Monte Carlo simulations). This is because the Gutzwiller ansatz provides a ``warm start" that aligns with the system’s topological and entanglement features, thereby narrowing the search space and mitigating the risk of local minima that arise from the complex, high-dimensional optimization in DMRG.

\subsection{SU(4) Kugel-Khomskii Model}

The SU(4) Kugel-Khomskii model extends the physics of spin-orbital interactions beyond the conventional SU(2) framework by incorporating orbital degeneracies~\cite{Kugel1982,Li1998}.
In this model, each lattice site hosts both spin and orbital degrees of freedom, and their interactions are described by a Hamiltonian that respects an enlarged SU(4) symmetry:
\begin{equation}
H = \frac{1}{2}\sum_{\langle i j \rangle} \left(  
\boldsymbol{\sigma}_i \cdot \boldsymbol{\sigma}_j +1 \right) \left(  \boldsymbol{\tau}_i \cdot \boldsymbol{\tau}_j +
	1 \right),  
\end{equation}
where $\langle ij\rangle$ denotes a nearest-neighbor bond. Here, $\boldsymbol{\sigma}$ and $\boldsymbol{\tau}$ represent Pauli matrices for spin and orbital degrees of freedom, respectively.
In particular, the equal footing spin and orbital degrees of freedom play a symmetric role as well and are promoted to SU(4) symmetry rather than the usual $\mathrm{SU(2)} \times \mathrm{SU(2)} \simeq \mathrm{SO(4)}$ symmetry at this fine-tuned point. 

Generally, the larger symmetry amplifies quantum fluctuations and might stabilize spin-orbital liquid ground states. Therefore, our Gutzwiller-guided DMRG method naturally provides a versatile tool for investigating possible emergent gauge fields and fractionalized excitations in the SU(4) Kugel-Khomskii model. 

We first briefly review the fermionic parton construction for SU(4) quantum magnets.
This starts by introducing four-flavor fermionic partons $f^\dagger_{i,m} \; (m=1,\ldots,4)$ and express the spin and orbital operators as
\begin{equation}
\begin{split}
&\sigma^a_{i} \rightarrow \bm{f}^\dagger_i\sigma^a\tau^0\bm{f}_i,\\
&\tau^b_{i}\rightarrow \bm{f}^\dagger_i\sigma^0\tau^b\bm{f}_i,\\
&(\sigma^a\tau^b)_i \rightarrow \bm{f}^\dagger_i\sigma^a\tau^b\bm{f}_i,    
\end{split}    
\end{equation}
where $\sigma^0$ ($\tau^0$) is the identity matrix in spin (orbital) space, and $\bm{f}^\dagger_i=(f^\dagger_{i,1},\cdots,f^\dagger_{i,4})$ are fermion creation operators. In order to restore the original Hilbert space, a single-occupancy constraint, $\sum_{m=1}^4 f^\dagger_{i,m}f_{i,m}=1$, has to be imposed at each site, which also defines the Gutzwiller projection for this SU(4) system.

\begin{figure}
    \centering
    \includegraphics[width=1.0\linewidth]{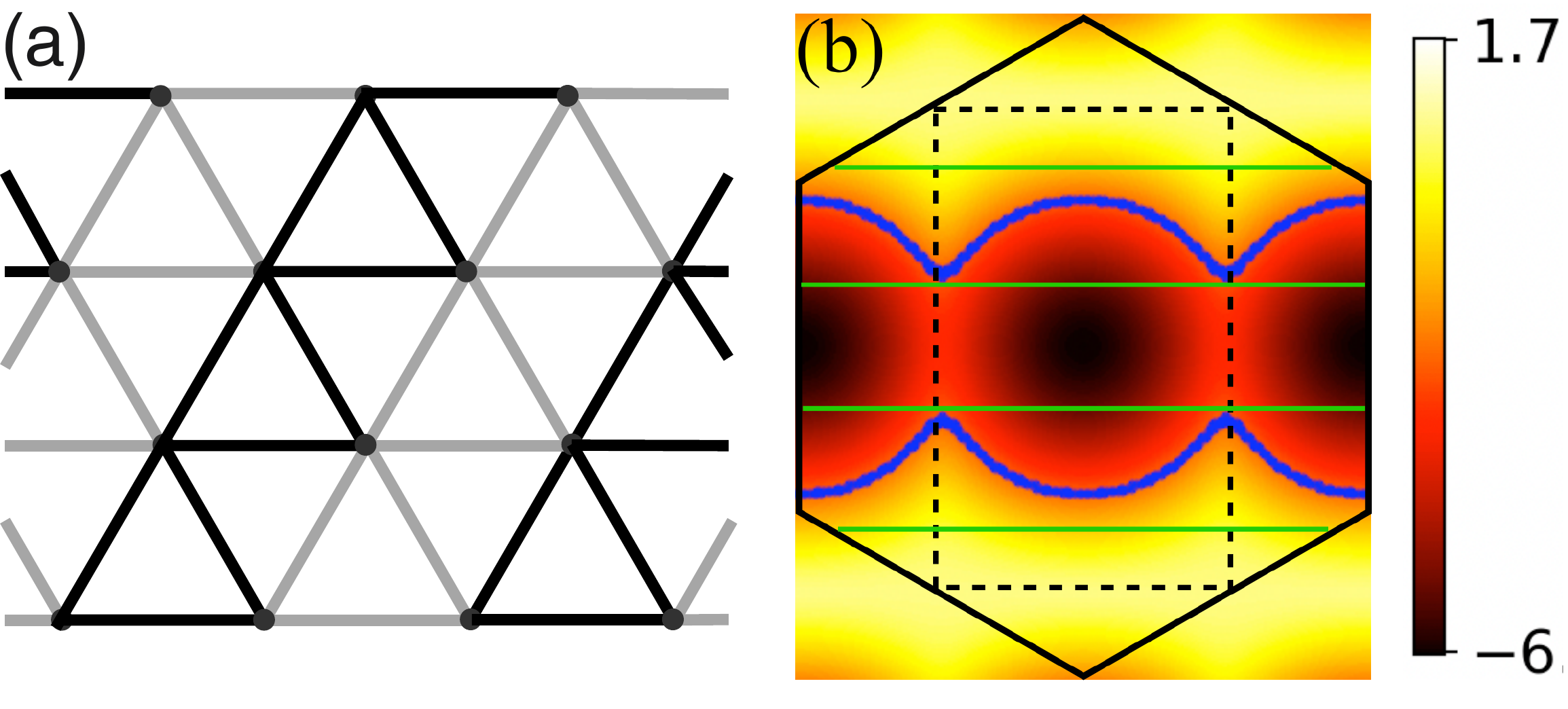}
    \caption{ (a) A sketch for the stripy parton state on a $L_y=4$ cylinder. The black and greay bonds represent $1+\delta$ and $1-\delta$ nearest neighbor bonds in Eq.~\eqref{eq:Hstripy}, respectively.  
    (b) The single-particle spectrum for the lower band of the stripy parton state with $\delta=0.15$ in the (unfolded) first Brillouin zone. The blue curve stands for the deformed Fermi surface at 1/4 filling, which consists of open orbits in the reciprocal space. Dashed lines enclose the folded Brillouin zone. Green lines represent momenta on XC4 cylinders that are allowed by the APBC along $y$ direction for partons. Figure from Ref.~\cite{Jin2021}.}
    \label{fig:stripy}
\end{figure}

With the help of the parton representation, a mean-field decomposition can be carried out to obtain various effective quadratic Hamiltonian of partons, which determines mean-field ground states and low-energy excitations. In the study of the SU(4) Kugel-Khomskii model on the triangular lattice, we consider a stripy state that breaks one of the lattice translational symmetries by doubling the unit cell, as shown in Fig.~\ref{fig:stripy}(a). Without loss of generality, we take $x$-direction as the symmetry breaking direction, and write down the mean-field Hamiltonian for such a stripy state as~\cite{Jin2022Bulletin}
\begin{equation}
	H_{\rm sp} =-\sum_{\langle ij \rangle,m} \left[1+(-1)^{r_i}\delta\right]\left( f^\dag_{i,m}f_{j,m} + f^\dag_{j,m}f_{i,m} \right), \label{eq:Hstripy}
\end{equation}
where $r_i$ is the column index of lattice site $i$, $r_{j}\ge{}r_{i}$ is assumed for nearest-neighbor bond $\langle{}ij\rangle$, and $\delta{}$ denotes the stripy strength.
A finite $\delta$ deforms the circular Fermi surface at $\delta=0$. We find that the deformed Fermi surface consists of open orbits in reciprocal space when $\delta>\delta_{c}\approx0.09$, as shown in Fig.~\ref{fig:stripy}(b).

This parton mean-field Hamiltonian~\eqref{eq:Hstripy}  gives rise to a Fermi sea ground state at quarter filling, whose Gutzwiller projected version can be converted into MPS with the MPO-MPS method~\cite{Wu2020}.
To perform MPS-related calculations on 2D lattices, we must must place the system on cylindrical geometries and work with finite-width cylinders.
Note that fermionic partons are coupled to an emergent $U(1)$ gauge field which can lead to a global gauge flux, $\Phi$, through the cylinder along the periodic boundary conditions. 
The time-reversal symmetry only allows $\Phi=0$ or $\Phi=\pi$, corresponding to periodic boundary condition (PBC) or anti-periodic boundary condition (APBC) for fermionic partons, respectively. 
Note that $\Phi$ is not a physical observable and is well-defined only with respect to a parton Hamiltonian with fixed $U(1)$ gauge. Therefore, choosing APBC along the $y$ direction for fermionic partons still yields PBC for spin and orbital degrees of freedom. 

For finite cylinders, choosing PBC and APBC for fermionic partons corresponds to different ways of cutting the Brillouin zone. 
Crucially, for a finite-width cylinder, the energy can be reduced by avoiding the cutting of momenta at any existing gapless points. Although it is difficult to prove this argument analytically, numerical validation is feasible using the Gutzwiller-guided DMRG method.

Using the combinative method, we obtain the ground states of the SU(4) Kugel-Khomskii model on a triangular lattice with different cylindrical boundary conditions and various system sizes. We then evaluate the variational ground-state energy of the stripy ansatz, and find that it reaches its minimum at $\delta \approx0.15$, indicating a stripy spin liquid state.
Moreover, we calculate the wave-function fidelity defined by
\begin{equation}
F = \langle\Psi_{\rm Gutzwiller}|\Psi_{{\rm DMRG}}\rangle.
\label{eq:fidelity} 
\end{equation}
Notably, $F$ is also maximized at $\delta{}\approx0.15$ with a value of $F\approx0.85$ on a $4\times8$ cylinder, in good agreement with the energy comparison. 

It is remarkable that the optimal value of $\delta$ does not open an energy gap on the parton Fermi surface. Thereby, a deformed parton Fermi surface consisting of open orbits is expected at quarter-filling in the two-dimensional limit. However, due to the global emergent U(1)-flux $\Phi$, none of the momenta allowed in these finite $L_y$ cylinders cuts the Fermi surface. For instance, $\Phi=\pi$ has a lower energy for $L_y=4$ cylinders (which leads to a gapped parton Fermi surface), while $\Phi=0$ is energetically favored for $L_y=6$ cylinders. Similar results have been found in the SU(4) Kugel-Khomskii model on the honeycomb lattice, which supports a Dirac spin liquid~\cite{Corboz2012,Jin2023}. On finite cylinders, the Dirac cones are also gapped by the emergent global U(1)-flux $\Phi$ under the same mechanism. 

The above results indicate another important advantage of the Gutzwiller-guided DMRG method, namely that it provides a diagnosis tool for analyzing the quality of Gutzwiller ansatz for a given Hamiltonian. 
As discussed above, the SU(4) Kugel-Khomskii model on the triangular and honeycomb lattices, while fundamentally gapless, can exhibit gapped features on finite cylindrical geometries, which might lead to misidentification as gapped states (if no special care has been taken). Consequently, one may misidentify the SU(4) Kugel-Khomskii model on quasi-1D geometries as a gapped QSL or other states when focusing only on local quantities like energetics and (short-distance) correlation functions. 
On the other hand, the Gutzwiller-guided DMRG method serves as a powerful tool for evaluating wave function fidelities $F$. Unlike local correlation functions, a high value of fidelity $F$, encoding non-local information, provides a more reliable indicator, effectively distinguishing true bulk behavior from finite-size artifacts.

\subsection{Kagome Heisenberg Antiferromagnet}

The $S=1/2$ kagome Heisenberg antiferromagnet (KHAF) is one of the most intriguing and challenging systems in the field of quantum magnetism, renowned for its potential to host exotic quantum spin liquids. The model describes a system of spin-$\frac{1}{2}$ particles arranged on a two-dimensional kagome lattice, where the spins interact via antiferromagnetic nearest-neighbor couplings. The Hamiltonian is given by
\begin{equation}
H_{\text{KHAF}} = \sum_{\langle ij \rangle_1} \mathbf{S}_i \cdot \mathbf{S}_j,\label{eq:KHAF_H}
\end{equation}
where $\langle ij \rangle_1$ denotes the 1st nearest-neighbor bonds. The geometric frustration inherent in the kagome lattice, combined with quantum fluctuations, suppresses conventional magnetic ordering. This makes the KHAF a prime candidate for realizing quantum spin liquid states.

Over decades, the ground state of the KHAF has been studied with many different approaches, but its nature is still under debate. Early theoretical studies suggested $\sqrt{3}\times{}\sqrt{3}$ magnetic order~\cite{Sachdev1992} or valence-bond crystal with spontaneously broken lattice symmetry~\cite{Mendel2007,Helton2007,lee2007,han2012,Norman16}. However, later numerical studies (in particular, DMRG and tensor networks) provided evidence that KHAF may instead host a quantum spin liquid ground state~\cite{Yan2011,Depenbrock2012,Jiang2008,jiang2012,nishimoto2013,Liao2017,Jahromi2020,He2017}. The distinction between gapped and gapless spin liquid phases has been a focal point of these studies.

With the Gutzwiller-guided DMRG method at hand, we have revisited the KHAF in Ref.~\cite{Sun2024}. It provides strong numerical evidence that the ground state of the KHAF might be a chiral spin liquid (CSL), with spontaneously broken time-reversal symmetry (TRS).
Previous works report that this Kalmeyer-Laughlin-type CSL is stabilized in an extended kagome Heisenberg model with the following Hamiltonian~\cite{Gong2014, Hu2015VMC}: 
\begin{equation}
H_{\text{ex}} = H_{\text{KHAF}} +  J'\left(\sum_{\langle ij \rangle_2} \mathbf{S}_i \cdot \mathbf{S}_j+\sum_{\langle ij \rangle_3} \mathbf{S}_i \cdot \mathbf{S}_j\right),
\end{equation}
where $\langle ij \rangle_n$ denotes the $n$-th nearest-neighbor bonds (the third-neighbor couplings only exist within the hexagons).
However, our results in Ref.~\cite{Sun2024} demonstrate that this CSL phase includes the $J'=0$ point, i.e., the KHAF with only 1st nearest-neighbor interactions.

\subsubsection{Chiral Spin Liquid Phase and Evidence}

A CSL is a gapped quantum spin liquid characterized by broken time-reversal as well as non-trivial topological order. In the parton representation, a Kalmeyer-Laughlin type CSL can be constructed using the following fermionic parton Hamiltonian~\cite{Ran2007}:
\begin{equation}
\label{eq:pt_mf}
	H_{\text{CSL}} = \sum_{\langle{}ij\rangle_1}\sum_{\sigma=\uparrow,\downarrow}(\chi_{ij}f^{\dagger}_{i\sigma}f_{j\sigma} + \text{H.c.}),
\end{equation}
where $\chi_{ij}=e^{\phi_{ij}}$ are link variables on the 1st nearest-neighbor bonds.
Due to SU(2) gauge redundancy~\cite{Wen02}, the ground states of Eq.~\eqref{eq:pt_mf} are distinguished by the gauge fluxes threading through elementary triangles and hexagons on the kagome lattice. The CSL state of particular interest corresponds to the ansatz with $\frac{\pi}{2}$-flux through triangles and zero-flux through hexagons.

Being a topologically ordered state, the degeneracy of the Gutzwiller-projected ground state, governed by $H_{\text{CSL}}$ in Eq.~\eqref{eq:pt_mf}, depends on the topology of the manifold. 
When placed on a cylinder, it supports four exact zero modes at the boundary, denoted as $d_{L\sigma}^{\dagger}$ and $d_{R\sigma}^{\dagger}$, where $L$ and $R$ represent the left and right boundary of the cylinder, respectively. Then, the minimally entangled states (anyon eigen basis)~\cite{zhang2012quasiparticle} of such a CSL are constructed as follows~\cite{tu2013b,Wu2020}:
\begin{equation}
    \label{eq:zeromodes}
    |\Psi_{1}\rangle = P_{G} d_{L\uparrow}^{\dagger}d_{L\downarrow}^{\dagger}|\Phi\rangle, \quad |\Psi_{2}\rangle = P_{G} d_{L\uparrow}^{\dagger}d_{R\downarrow}^{\dagger} |\Phi\rangle~,
\end{equation}
where $|\Phi\rangle$ is a Fermi sea state with all negative energy modes of Eq.~\eqref{eq:pt_mf} being occupied, and $P_G$ is the Gutzwiller projector ensuring single occupancy. Using the MPO-MPS method~\cite{Wu2020}, the CSL anyon eigenbasis in Eq.~\eqref{eq:zeromodes} can be efficiently converted into MPSs to carry out the Gutzwiller-guided DMRG process.

In the KHAF, we have identified the CSL phase by several key signatures:

1. \textbf{Spontaneous TRS breaking at $J'=0$}: 
We perform DMRG calculations initialized with randomly generated MPSs (denoted as ``Random-DMRG'' below) as well as the Gutzwiller-guided DMRG initialized with $|\Psi_1\rangle$. The ground-state energies obtained by Random-DMRG and Gutzwiller-guided DMRG are almost identical, and the ground-state energy per site is consistent with the best available DMRG results~\cite{Yan2011,Depenbrock2012}. With different initializations, the ground state $|\Psi_{\text{I}}\rangle$ and its time-reversal partner $|{\Psi_{\text{I}}}^{*}\rangle$ are obtained at $J'=0$, indicating the breaking of TRS. 

The spin chirality operator, which is defined on elementary triangles $\triangle$ of sites $(i,j,k)$ as   
\begin{equation}
\chi_{ijk} = \mathbf{S}_i \cdot (\mathbf{S}_j \times \mathbf{S}_k),
\end{equation}
serves as a marker of TRS breaking~\cite{Wen89}. However, directly evaluating the spin chirality order is challenging in previous DMRG studies that predominantly employed real-number MPS. Because the spin chirality operator transforms as $\chi_{ijk} \rightarrow -\chi_{ijk}$ under time reversal, the expectation value of $\chi_{ijk}$ with respect to a real-number MPS is always zero.
Instead, these previous works calculated the chiral-chiral correlation function, resulting in an expectation value of spin chirality that is too small to be distinguished from (numerical) zero.
On the other hand, our Gutzwiller-guide DMRG can target the two-fold Kramers degeneracy, which allowed us to directly evaluate the local spin chirality $\langle\chi_{ijk}\rangle$. 
We find that $\langle\chi_{ijk}\rangle\sim{}10^{-4}$ is small but does not vanish numerically (note that the truncation error in our DMRG calculations is $10^{-6}$), and is stable against finite-size scaling.

2. \textbf{Topological Degeneracy:} Another essential signal of the CSL at $J' = 0$ is the existence of (quasi-)degenerate ground states, $|\Psi_{S}\rangle$, in the semion sector. For the CSL state at $J'=0.2$, $|\Psi_{S}\rangle$ can be clearly established, as the characteristic level counting of the entanglement spectra is consistent with the chiral SU(2)$_1$ Wess-Zumino-Witten conformal field theory. 
We further found that $|\Psi_{S}\rangle$ can be adiabatically evolved from $J' = 0.2$ to $J' = 0$ using the Gutzwiller-guided DMRG method, in which the wave function fidelity $F$ between two neighboring-$J'$ states, e.g. $|\Psi_{S}(J')\rangle$ and $|\Psi_{S}(J'-0.01)\rangle$, is always close to unity for $0.0\leq J' \leq 0.2$. The entanglement spectra of $|\Psi_{S}\rangle$ also show qualitative consistency in the region of $0 \leq J' \leq 0.2$. Moreover, we showed that the per-site energy variances of $|\Psi_{S}\rangle$ for the KHAF model at $J'=0$, $\sigma^{2}/N$ is as small as $\sim 10^{-4}$, suggesting that $|\Psi_{S}\rangle$ is indeed an eigenstate of the Hamiltonian in Eq.~\eqref{eq:KHAF_H}. 

The identification of the CSL phase in the KHAF is robust against finite-size scaling. For instance, by pushing the bond dimension up to $D=18000$ and the circumference of cylinder up to $L_y=8$, we confirmed that the results on spin chirality and semion sectors are more evident as $D$ and $L_y$ increase. 
Nevertheless, challenges remain in fully characterizing the CSL phase in the thermodynamic limit. The finite circumference of the cylinder geometry limits the resolution of long-range correlations, and the growth of the computational cost with bond dimension imposes practical constraints. Future studies employing wider cylinders or alternative tensor network approaches, such as projected entangled pair states, may help address these limitations.

\section{Summary and outlook}

In this review, we summarize recent progress on integration of Gutzwiller-projected wave functions with the DMRG method. By combining the strengths of Gutzwiller projected fermionic parton wave functions and MPS techniques, we have devised a promising hybrid approach for investigating 2D strongly correlated systems. The key advantages are: 1) the notorious ``local minimum'' issue in DMRG can be circumvented, and the precision of DMRG can be improved by orders of magnitude without extra computational cost; 2) apart from local correlation functions, the new method can be used to compute more essential non-local observables such as wave function fidelity, which allows us to directly diagnose the quality of the parton wave functions; 3) it provides a versatile tool to target (topologically) degenerate ground states in different (topological) sectors.

A significant result of our investigation has been the accurate capture of complex phenomena such as topological order, fractionalized excitations, and the subtle interplay between competing interactions. For the cases of the Kitaev honeycomb model, the SU(4) Kugel-Khomskii model, and the KHAF model, the hybrid approach allows us to provide deeper insights into emergent behaviors in these systems. The flexibility of the method in adapting to various Hamiltonians and lattice geometries demonstrates its potential as a universal tool in the study of quantum spin liquids and other strongly correlated phases of matter.

The combination of Gutzwiller-projected wave functions with DMRG represents a significant step forward in the numerical study of many-body physics, providing a pathway to systematically explore and characterize quantum phases that emerge from strong electron correlations. This hybrid approach is expected to play a crucial role in resolving future challenges in the field of quantum magnetism and beyond.

Along this direction, there are several key issues that warrant further investigation:

1. Developing more efficient algorithm for converting fermionic Gaussian states into MPSs. The existing method introduced in Ref.~\cite{Wu2020} is effective yet somewhat brute-force and might not constitute the best strategy, especially within 2D systems. A potential alternative approach is to divide the entire system into several separate subsystems as small as possible. In this context, the Wannier orbitals correspond precisely to the Schmidt vectors of the reduced density matrices for each subsystem, which can be easily computed using the fermionic Gaussian state theory.

2. The transformation of the Gutzwiller-projected parton state into a 2D tensor network state, such as projected entangled pair state (PEPS), holds great promise. Compared to MPS, the optimization of PEPS relies even more heavily on the choice of initial states.
Considering PEPS can capture entanglement entropy beyond the area law, this transformation could potentially unlock new ways of understanding and simulating complex quantum systems~\cite{Yang2023,Li2023}.

3. The Gutzwiller projection is a powerful tool, but its application has been mainly restricted to the case of an infinitely large Hubbard $U\rightarrow\infty$.
However, the effects of charge fluctuations are inherently unavoidable in realistic systems.
Extending this framework to accommodate finite Hubbard $U$ interactions~\cite{Zhou2024} and/or deviations from half-filling is a pressing issue for future research.

4. The Gutzwiller-guided DMRG has been demonstrated to be highly efficient in targeting the ground state in specific topological sectors. Building on this, the next step involves identifying potential applications for a range of topological systems, including but not limited to symmetry-protected topological states~\cite{cai2025} and symmetry-enriched topological states.
\begin{acknowledgments}

We are very grateful to the many colleagues (too many to list here) who have collaborated with us and contributed to the development and application of the Gutzwiller-guided DMRG method. 

\textbf{Funding --} Y.Z. is supported in part by the National Key Research and Development Program of China (Grant No. 2022YFA1403403), the National Natural Science Foundation of China (Grants No. 12274441 and No. 12034004).
H.-K.J. acknowledges the support from the start-up funding from ShanghaiTech University.

\textbf{Competing Interest --} The authors declare that they have no competing interest.

\textbf{Author Contribution --}  All authors discussed the results and contributed to writing the article.

\textbf{Availability of Data and Materials --} This review article compiles data and materials that are all available in published papers or pre-prints in arXiv.

\end{acknowledgments}

\bibliography{merged}

\end{document}